# Evaluasi Celah Keamanan Web Server pada LPSE Kota Palembang

Muhammad Ilham Daniel[1], Leon Andretti Abdillah[2], Kiky Rizky Nova Wardani[3]
[1]Program Studi Teknik Informatika, Fakultas Ilmu Komputer, Universitasa Bina Darma
[2,3]Program Studi Sistem Informasi, Fakultas Ilmu Komputer, Universitasa Bina Darma
Palembang, Indonesia
[1]m.ilhamdaniel@ymail.com, [2]leon.abdillah@yahoo.com

**Abstract.** Along the development of information technology systems among the public at large, also develops information systems to facilitate the public to access and search for information in the form of a website. Electronic Procurement Service (LPSE) Palembang is a business unit set up to organize the service system of government procurement of goods or services electronically. And to allow companies or providers that want to follow the procurement of goods or services, LPSE providing a website that can be accessed from anywhere so the company or provider to follow the procurement of goods or services without having to come to the office LPSE. In the management of its website, LPSE Palembang has its own web server so that the need to consider the existing security system on the web server. Web servers often become the target of attacks by an attacker. This study is set to test the security system of the web server to find out if a web server is secure or not of the crime committed by an attacker. This research involves penetration testing with multiple applications. The results show some holes and suggestions.

## 1 Pendahuluan

Seiring semakin berkembangnya teknologi informasi (TI) dikalangan masyarakat luas, berkembang pula sistem informasi (SI) yang dapat memudahkan masyarakat untuk mengakses dan mencari informasi dari media *website*. *Website* merupakan layanan sistem informasi yang dapat diakses melalui jaringan *internet* oleh pengguna diseluruh dunia. Pada umumnya *website* memiliki empat elemen dasar [1], yaitu: 1) *browser*, 2) *server*, 3) *uniform resource locator* (URL), dan 4) *pages*. Web server merupakan sebuah perangkat lunak *server* yang berfungsi menerima permintaan *yyper text transfer protocol* (HTTP) atau *hyper text transfer protocol secure* (HTTPS) dari klien yang dikenal dengan *web browser* dan mengirimkan kembali hasilnya dalam bentuk *website* (halaman *web*) yang pada umumnya berbentuk dokumen *hyper text markup language* (HTML).

Layanan Pengadaan Secara Elektronik (LPSE) Kota Palembang merupakan unit kerja yang dibentuk untuk menyelenggarakan sistem pelayanan pengadaan barang/jasa pemerintah secara elektronik (http://lpse.palembang.go.id/eproc/). Dan untuk memudahkan perusahaan/penyedia yang ingin mengikuti pengadaan barang/jasa, LPSE menyediakan situs website dengan domain http://www.lpse.palembang.go.id/ yang dapat diakses dari manapun sehingga perusahaan/penyedia tetap dapat mengikuti pengadaan barang/jasa tanpa harus datang kekantor LPSE. Dalam pengelolaan website LPSE, LPSE Kota Palembang memiliki web server sendiri sehingga perlu memperhatikan sistem keamanan yang ada pada web server tersebut. Karena web server seringkali menjadi target serangan yang dilakukan oleh seorang attacker. Hal ini dapat terjadi karena adanya celah keamanan pada web server yang dapat dimanfaatkan penyerang untuk dapat masuk kedalam server. Walaupun pada umumnya serangan yang terjadi hanya menimbulkan kesan negatif dan memalukan seperti mengganti halaman website (defacing) bukan tidak mungkin penyerang dapat membuat kerusakan yang lebih parah dan sangat merugikan.

Dalam pengelolaan web server LPSE masalah yang terjadi saat ini yaitu sistem yang sering mati dan belum pernah dilakukannya pengujian terhadap sistem keamanan pada web server LPSE Kota Palembang. Pengujian sangatlah penting untuk mengetahui apakah web server sudah aman atau belum dari tindak kejahatan yang dilakukan oleh seorang *attacker* [1]. Oleh sebab itu, perlu dilakukan pengujian pada web server LPSE Kota Palembang. Dan dari hasil pengujian yang dilakukan peneliti dengan melakukan test error dan scanning pada web server LPSE menggunakan beberapa aplikasi ditemukan beberapa port server yang terbuka, terlihat host yang digunakan dan juga terdapat celah keamanan lainnya yang rentan.

Berdasarkan permasalahan di atas untuk menjaga keamanan webserver LPSE, peneliti melakukan evaluasi pada web server LPSE Kota Palembang sehingga jika menemukan celah keamanan pada web server dapat melaporkannya kepada pihak LPSE dan segera diperbaiki.

Tujuan dari penelitian ini adalah: 1) Mengevaluasi keamanan eksternal sistem pada web server LPSE Kota Palembang, 2) Diharapkan dapat membantu administrator LPSE dalam mengidentifikasi celah keamanan yang ada pada web server sehingga dapat segera menutup dan memperbaiki celah keamanan tersebut.

Agar penelitian tetap terarah dan tidak terlalu menyimpang dari permasalahan yang ada, maka batasan masalah pada penelitian ini adalah : 1) Pengujian dilakukan pada web server LPSE Kota Palembang yaitu melalui situs www.lpse.palembang.go.id., 2) Pengujian dengan pendekatan white box testing, 3) Melakukan beberapa perbaikan bersama dengan admin LPSE Kota Palembang, dan 4) Peneliti tidak melakukan perbaikan terhadap celah keamanan yang ada pada aplikasi website.

Adapun manfaat dari penelitian ini adalah : 1) Meningkatkan keamanan pada web server LPSE Kota Palembang., 2) Bagi administrator LPSE penelitian ini bermanfaat untuk mengidentifikasi celah keamanan yang ada pada web server sehingga dapat langsung menutup dan memperbaiki celah keamanan tersebut, dan 3) Bagi pengguna penelitian ini bermanfaat untuk keamanan dan kenyamanan dalam mengakses situs LPSE.

Sejumlah literatur telah penulis pelajari untuk mendukung pelaksanaan penelitian ini, antara lain: 1) Ismail [2] meneliti keamanan server web dengan alamat website http://www.beta.kotimkab.go.id. Penelitian ini melakukan *scanning port* dengan aplikasi *nmap*, dan teknik *SQL Injection*. Hasilnya terdapat beberapa celah keamanan pada modul Agenda dan pada web scanning menemukan beberapa berkas yang usang. Sehingga server web dan aplikasi didalamnya membutuhkan perawatan dan penanganan khusus dengan melakukan *audit*, *upgrade software* dan *hardware* secara berkala, sehingga pengembangan situs web dan layanan internet didalam website pemerintahan menjadi lebih aman dari gangguan peretas di dunia maya, 2) Metasari [3] melalukan Analisis Keamanan Website di Universitas Muhammadiyah Surakartadengan menggunakan *software Acunetic website vulnerability scanner*. Adapun berbagai *web alerts* yang berhasil ditemukan berupa *SQL Injection*, *Cross Site Scripting*, dan berbagai *web alerts* lainnya, 2) Siagian [4] melakukan analisis "Vulnerability Assessment pada Web Server Universitas Bina Darma". Peneliti menggunakan beberapa aplikasi yaitu *Acunetic*, *Nikto*, *OpenVAS*, dan *Retina Web Scanner*. Dan dari hasil uji coba menggunakan aplikasi-aplikasi tersebut didapat beberapa kerentanan pada *web server* Universitas Bina Darma yang diakibatkan dari aplikasi yang di-*install* di *server* kadaluwarsa.

## 2   Metode Penelitian

Metode yang digunakan dalam penelitian ini menggunakan penelitian tindakan atau *action research* [5]. Penelitian *Action Research* dipilih karena pada penelitian ini langsung tertuju pada objek penelitian yaitu mengevaluasi celah keamanan (*vulnerability*) pada *web server* LPSE Kota Palembang. Tahapan yang dilakukan adalah sebagai berikut: 1) Mendiagnosa(*diagnosing*), 2) Melakukan perencanaan tindakan (*action planning*), 3) Melakukan evaluasi (*evaluating*), dan 4) Menentukan pembelajaran dari hasil penelitian (*Learning*).

Peneliti melakukan diagnosa celah keamanan pada *web server* dengan teknik *penetration testing* [6] yang terdiri dari tahapan: 1) *planning*, 2) *discovery*, dan 3) *attack*. Selanjutnya, peneliti akan melakukan analisis terhadap hasil diagnosa, dan melakukan beberapa perbaikan bersama dengan admin LPSE dan juga peneliti akan melakukan pengujian ulang pada celah keamanan yang telah diperbaiki.

## 3   Hasil dan Pembahasan

Dari hasil *diagnosing* yang telah dilakukan, peneliti menemukan celah keamanan pada *web server*. pada saat melakukan *test error* halaman web menampilkan pesan *error* yang berisi informasi sensitif berupa aplikasi *web server* yang digunakan yaitu apache 2.2.16 (Debian) yang dapat dijadikan langkah awal *attacker* untuk menyerang.

### 3.1  Melakukan Tindakan (*Action Taking*)

Pada action taking peneliti melakukan analisis terhadap celah keamanan  yang telah ditemukan, kemudian memperbaiki beberapa celah keamanan yang ditemukan bersama dengan admin LPSE, dan selanjutnya melakukan pengujian ulang.

Student Colloquium Sistem Informasi & Teknik Informatika (SC-SITI)
Palembang, 21-22 Agustus 2015Peneliti melakukan analisis terhadap celah keamanan yang ditemukan dari hasil diagnosa yang dijelaskan pada tabel 1.

**Table 1.** Celah Keamanan *Web Server*

| No | Celah Keamanan | Dampak dari Celah Keamanan |
|---|---|---|
| 1. | *Test Error* | Dapat menampilkan pesan error pada halaman web berisi informasi sensitif berupa aplikasi web server. |
| 2. | *Port Terbuka* | Terdapat 7 ports yang terbuka pada web server seperti contoh port 22 service ssh yang dapat diakses oleh orang luar melalui internet |
| 3. | *Blind SQL Injection* | Dapat mengambil isi dari database LPSE seperti user id dan password user penyedia. |

Setelah dikethui sejumlah celah keamana, maka dilakukanlah perbaikan. Perbaikan yang dilakukan peneliti yaitu pada celah keamanan test error dan pada sejumlah ports yang terbuka, antara lain dengan cara: 1) Melakukan perbaikan *test error* (gambar 1), dan 2) Melakukan *filter port* (gambar 2).

Perbaikan *test error* dilakukan oleh peneliti bersama admin LPSE melakukan perbaikan dengan tidak menampilkan aplikasi web server pada saat melakukan *test error* pada *website* LPSE dengan cara menambahkan perintah " Server signature Off " kedalam file /etc/apache/apache2.conf pada *web server*. Kemudian peneliti dan admin LPSE menutup dan mem-filter sejumlah *ports* yang terbuka pada *web server* LPSE Kota Palembang.

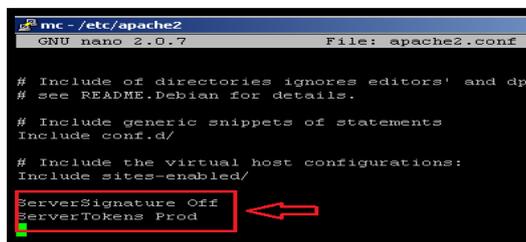

**Gambar 1.** Melakukan perbaikan *test error*

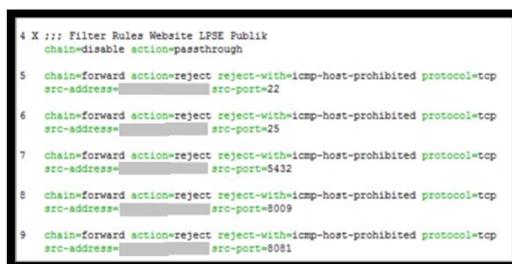

**Gambar 2.** *Filter ports*

### 3.1 Melakukan Pengujian Ulang

Peneliti kembali melakukan pengujian *test error* pada web LPSE. Hasilnya setelah dilakukan perbaikan web tidak menampilkan info aplikasi *web server* yang digunakan dan hanya menampilkan halaman yang berisi pesan error 404 (gambar 3).

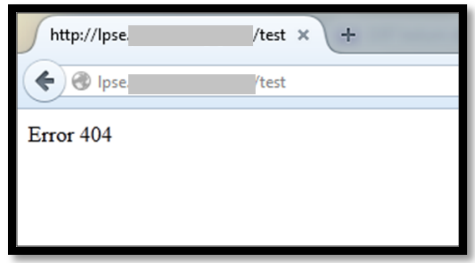

**Gambar 3.** Hasil *test error*

Selanjutnya meneliti melakukan scanning ulang menggunakan aplikasi nmap. Hasilnya sejumlah *ports* yang terbuka sudah berhasil di-*filter*. Dibawah ini merupakan rincian penjelasan port - port yang sudah berhasil difilter.

**Table 2.** *Filter Ports*

| Port | | State toogle closed [0] filtered [5] | Service |
|---|---|---|---|
| 22 | Tcp | Filtered | Ssh |
| 25 | Tcp | Filtered | Smtp |
| 80 | Tcp | Open | http |
| 5432 | Tcp | Filtered | Postgresql |
| 8009 | Tcp | Filtered | ajp13 |
| 8080 | Tcp | Open | http |
| 8081 | Tcp | Filtered | Blackice-icecap |

### 3.2 Evaluasi (*Evaluating*)

Pada pembahasan peneliti melakukan beberapa perbaikan celah keamanan bersama dengan admin LPSE Kota Palembang dan juga melakukan pengujian ulang terhadap celah keamanan yang telah diperbaiki. Setelah dilakukan perbaikan sejumlah *ports* yang terbuka sudah berhasil ditutup dan di *filter*. Dan juga pada saat dilakukan *test error* halaman *web* sudah tidak menampilkan pesan *error* yang berisi informasi aplikasi *web server* yang digunakan.

### 3.3 Pembelajaran (*Learning*)

Tahapan ini merupakan bagian akhir untuk mendapatkan kesimpulan dan saran dari evaluasi celah keamanan yang telah dilakukan pada LPSE Kota Palembang. Tahapan ini dijelaskan lebih rinci pada bagian akhir Kesimpulan dan Saran.

## 4 Kesimpulan

Berdasarkan hasil analisis dan ujicoba terhadap celah keamanan web serve pada LPSE, maka dapat disimpulkan:
1. Masih terdapat celah keamanan pada web server LPSE.
2. Perbaikan yang dilakukan hanya pada beberapa celah keamanan.
3. Perlu dilakukan pengujian untuk memeriksa kerentanan yang ada pada web server.
4. Sara yang dapat penulis berikan, antara lain: a) Perlu dilakukan evaluasi terhadap celah keamanan web server secara berkala, dan b) Melakukan *update* aplikasi-aplikasi yang sudah kadaluwarsa.

## Daftar Pustaka